\documentclass[USenglish,twocolumn]{article}

\usepackage[utf8]{inputenc}
\usepackage[big]{dgruyter}
\usepackage{float}
\usepackage{natbib}

\begin{document}

  \articletype{Research Article{\hfill}Open Access}

  \author*[1]{Thomas M. Boudreaux}

  \affil[1]{Department of Physics, High Point University, One University Parkway, High Point, NC 27268 USA, E-mail: thomas@boudreauxmail.com}

  \title{\huge The applications of deep neural networks to sdBV classification}

  \runningtitle{Deep neural networks classifying sdBV stars.}


  \begin{abstract}
	{With several new large-scale surveys on the horizon, including LSST, TESS, ZTF, and Evryscope, faster and more accurate analysis methods will be required to adequately process the enormous amount of data produced. Deep learning, used in industry for years now, allows for advanced feature detection in minimally prepared datasets at very high speeds; however, despite the advantages of this method, its application to astrophysics has not yet been extensively explored. This dearth may be due to a lack of training data available to researchers.  Here we generate synthetic data loosely mimicking the properties of acoustic mode pulsating stars and we show that two separate paradigms of deep learning -- the Artificial Neural Network And the Convolutional Neural Network -- can both be used to classify this synthetic data effectively. And that additionally this classification can be performed at relatively high levels of accuracy with minimal time spent adjusting network hyperparameters.}
\end{abstract}
  \keywords{Deep Learning, sdBV}

  \journalname{Open Astronomy}
  \startpage{1}
  \received{09/29/17}
  \revised{11/7/17}
  \accepted{11/30/17}

  \journalyear{2017}
  \journalvolume{1}

\maketitle
\section{Introduction}
The amount of data products produced by researchers has ballooned over the last 20 years, and with surveys such as The Large Synoptic Survey Telescope (LSST) expected to produce terabytes of data per night \citep{LSST09} it is clear that fast data analysis methods are a necessity. However, even without these next generation surveys there is already more data extant than can be effectively dealt with via the most common analysis procedures. We can see this problem highlighted in a recently identified contact binary showing an orbital period decay so extreme that in 2022 the system will experience a nova \citep{Mol17}. Despite this system manifesting a $\dot{P}$ visible within the time domain, a case study on this finding was only recently published. The time delay between observation and findings is often due to extensive amounts of data produced, and this delay highlights the data problem facing the astronomy community: as more and more data become available, interesting systems, even those with high signal--to--noise ratios (S/N), will often be buried below mounds of mundane targets. 

However, methods do exist to make data analysis more efficient. Deep learning -- the general term for a set of machine learning algorithms loosely inspired by the structure of biological brains -- is one such a method; it allows for feature detection in minimally prepared datasets. This last point, allowing minimally prepared data to be used, is key, as it opens the door for nearly raw data to be used in analysis, drastically reducing the time between when an observation is taken, and when a discovery is made. Deep Learning thus significantly reduces the search costs associated with astronomical discovery.

Investigations into the applications of deep learning to astrophysics are still in their infancy. Previous work includes analysis of aLIGO data \citep{Geo16}, galactic morphology classification \citep{Hue15}, and asteroseismological classification of red giant branch stars \citep{Hon17}, among others. Here we present preliminary results of our use of deep learning to analyze synthetic photometry of hot subdwarf B (sdB) stars and classify them as rapidly--pulsating sdB (sdBV$_r$) stars or not observed to vary (NOV) stars. sdB stars are extreme horizontal branch objects believed to have formed from red giants that lost their outer H envelopes while ascending the red giant branch, likely due to interactions with a nearby companion \citep{Heb16}. For further detail on the formation, properties, and pulsations of sdB stars see \citet{Heb16}.

We investigate the effectiveness of both traditional feed-forward artificial neural networks (ANNs)\citep{Sch15}  and feed-forward convolutional neural networks (CNNs)\citep{Sch15} in the {\em binary} classification of sdBV$_{r}$. Importantly, we only aim to classify a target as either ``pulsating'' or ``not observed to vary''. No attempt is made here at feature (such as pulsation amplitude/frequency) extraction. We use the Python model Keras \citep{Cho15} with the Tensor-flow \citep{Mar15} backend to generate, train, and validate all models presented in this paper.

Neural networks must be trained, and this training requires a large amount of already classified data. We develop a Python module -- astroSynth \citep{Astro17} -- to produce synthetic light curves whose noise properties mimic those seen in real data. astroSynth is used to produce 100,000 light curves. We then use 80 percent of these to train an ANN and the remaining 20 percent to validate the ANN's performance. Finally another function of astroSynths generates 100,000 ``virtual targets'' -- that is non-continuous light curves, to be analyzed with CNNs. 

We find that with minimal tuning of network structure we can achieve $\sim90$ percent accuracy in classification down to a S/N of 3.44 using the ANN and $\sim90$ percent accuracy down to S/N of 1.56 with the CNN. While these results are promising, and could most likely be improved upon by tuning the hyperparameters of the network, we elect not to do this, both because it is beyond the scope of this paper, and because we anticipate moving away from ANNs and CNNs in the future due to some fundamental constraints of feed-forward networks. Instead, we hope to focus future work on the use of Recurrent Neural networks (RNNs) \citep{Sch15}  which are better suited for work with time series data such as we have.

\section{Deep Learning}
Despite deep learning's wide-spread adoption in industry, including heavy use by firms such as Google, Facebook, Twitter, and Tesla, adoption of these algorithms has thus far been quite limited in astronomy. In the following sections, we provide a quick overview of the basic structure and principles that underlie the two network paradigms under investigation (ANNs and CNNs).

\subsection{Artificial Neural Networks} \label{ANNsSec}
\begin{figure}
	\centering
    \includegraphics[scale=0.5]{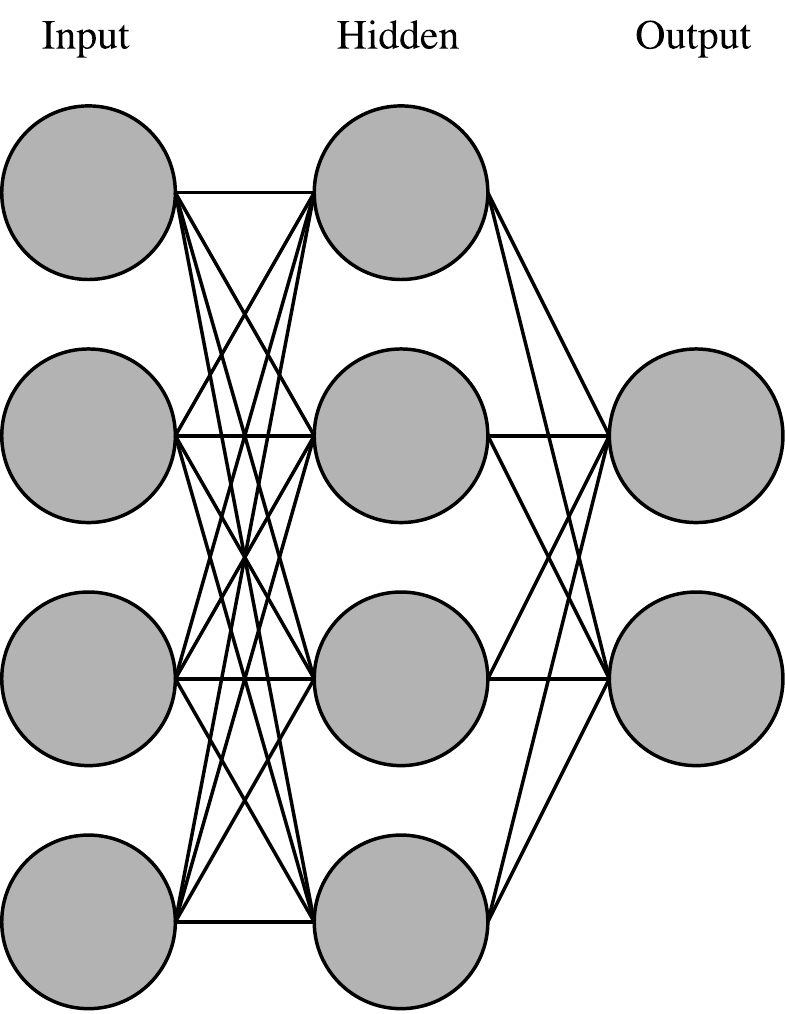}
    \caption{Characteristic Structure of an Artificial Neural Network (ANN). This network shows an input layer of dimension four, therefore the network expects a four vector, one hidden layer, and a two vector output. Typical production networks will have more and larger hidden layers. Note that each neuron (the grey circles) in each layer is connected to each neuron in the next layer.}
    \label{ANN}
\end{figure}

An evolution of the perceptron \citep{Ros58}, the artificial neural network (ANN) was an early kind of neural network to gain widespread usage. It arose with the discovery that stacking multiple layers of perceptrons can create a structure that is very efficient at modeling functions. Due to its stacked, sequential nature, an ANN is referred to as a feed-forward neural network. Each layer of an ANN is composed of cells that sum all incident inputs, and apply some non-linear function to the result of that summation. These cells are called neurons. Each neuron in a layer is connected to every neuron in the next layer (Figure \ref{ANN}). Consequently, these kinds of layers are known as ``fully-connected layers.'' Further connections between neurons, called synapses, should be thought of as weights assigning importance to different features that the network has extracted. Therefore, each connection can be imagined as the product of some weight and whatever values pass along it. 

More formally, a network's inputs, $\mathbf{x} = \mathbf{z}_{0}$, are passed forward through the network from some layer $n$ to the following layer $n+1$ via equation

$$\mathbf{z_{n+1}} = A(W_{n+1}\mathbf{z_{n}}+\mathbf{b_{n+1}})$$
\noindent where $z$ is the output from each layer, $W$ is a weight matrix, $\mathbf{b}$ is a bias vector, and $A$ is a non-linear activation function. Common activation functions include the logistic function, hyperbolic tangent, and rectified linear units (ReLU). It is also common to inject dropout layers --- which essentially throw away the inputs from a certain percentage of incident cells in order to limit over--fitting of data --- in-between fully connected layers. The output of the final layer  ($\hat{y}$) is used as the output of the network as a whole.

\begin{figure}
	\centering
	\includegraphics[scale=0.25]{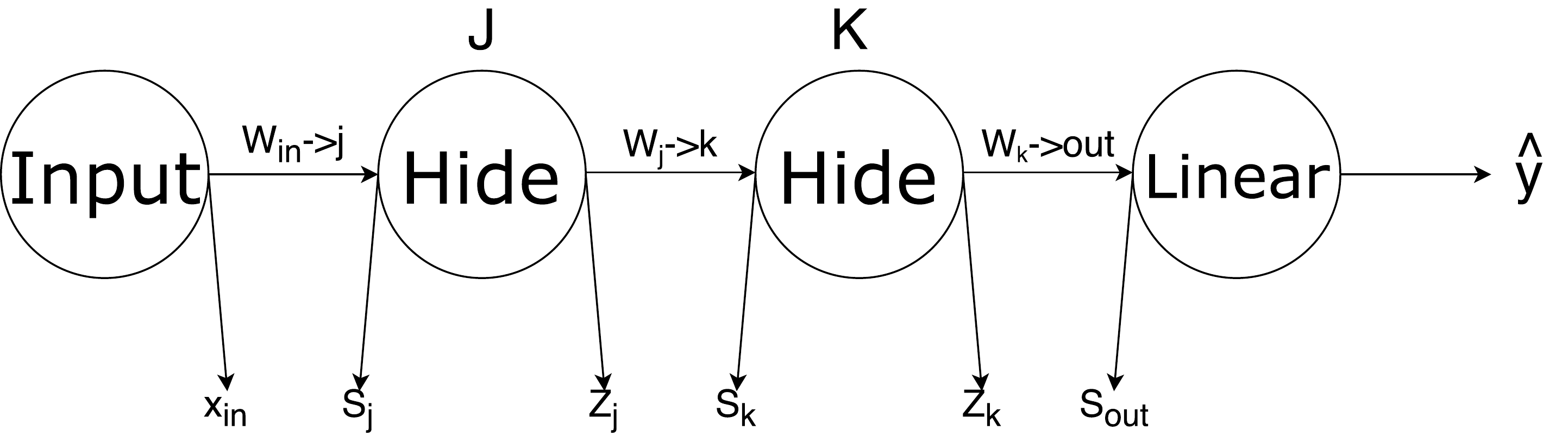}
	\caption{General mathematical structure of an ANN. Inputs, $\mathbf{x}$, are passed into the Network at the input neurons. The weight matrix $W_{i}$ associated with the input layer $I$ is multiplied by $\mathbf{x}$ yielding $s_{j}$, with the result of that operation then activated using some function yielding $z_{j}$. The same process is repeated to move to layer $K$. Finally $W_{k}$ is multiplied by $K$, activated with a {\em linear} activation function (often a softmax) and used as the network output.}
	\label{MANN}
\end{figure}

Another method of visualizing an ANN can be seen in Figure \ref{MANN}. It is important to note that ANNs take an input vector of a {\em predefined} size, and return an output vector of a predefined size. In the event of a data set whose elements are of variable size, an ANN will either be of limited use, or steps will have to be taken to account for the size difference in data elements. 

When a network is first instantiated the weight matrices are randomly set; therefore, for the output to provide insight into one's data the weights must be tuned. This process is called training. Deep learning falls into the category of supervised learning (this is as opposed to unsupervised learning algorithms such as K-Means Clustering) where in order to train a network the expected output values are required. Training the network begins by comparing the network output to the expected output, and computing the absolute error. A process known as back propagation then allows for that error to be carried back along the network, determining what portion of the error is due to each layer as it goes. The weights of the layers are then slightly  adjusted (limited by a user-defined learning rate, $\eta$) in the direction of reduced output error based on how much each layer contributes to the overall error. Networks often need to be trained on a large amount of data in order to produce reliable results. As $\eta$ is kept low in an attempt to avoid over-fitting the training-data set, back-propagation is the slowest part of ANN usage with the actual amount of time required to train being heavily dependent not only on the total amount of data but also the complexity of the network structure. Once the network is trained it can be used for its intended purpose, or retrained if new data becomes available.

\subsection{Convolutional Neural Networks}
\begin{figure}
	\centering
    \includegraphics[width=0.5\textwidth]{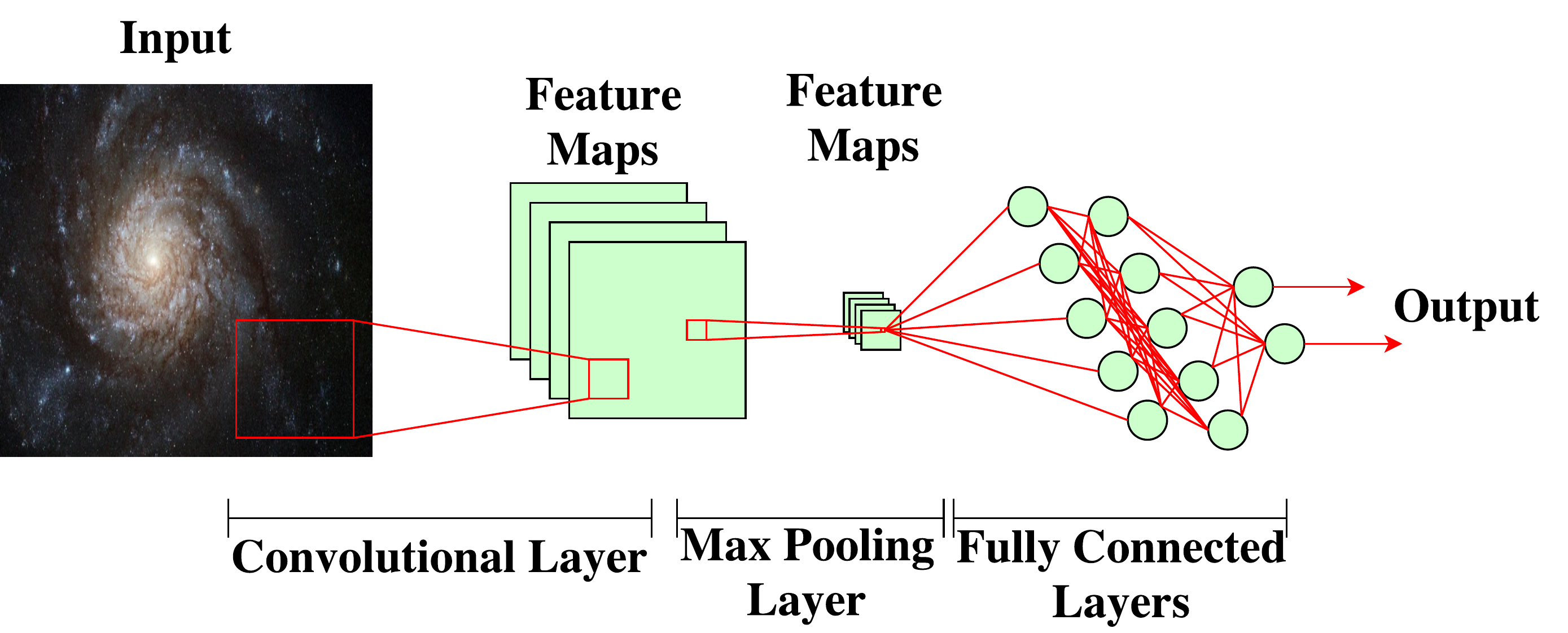}
    \caption{General Structure of a Convolutional Neural Network showing one convolutional layer (CNN), one pooling layer, a set of fully connected layers, and an output layer. Production CNNs traditionally have much more complex structures. Note that Pooling layers will almost always follow convolutional layers. Also note that there is some mechanism (often referred to as a flattening layer) to convert the 2D output of the pooling layer to the one dimensional input the fully connected layers expect. This mechanism is not shown here.}
    \label{CNN}
\end{figure}

Heavily inspired by the biological structures underlying vision \citep{Sch15} , convolutional neural networks (CNN) have proven extremely effective in image classification problems and have accordingly been widely adopted in recent years. CNNs classically take input of two dimensional data (however CNNs in both higher and lower dimensional space do exist), then pass data through convolution, pooling, and traditional fully connected layers among others (Figure \ref{CNN}). CNNs, like ANNs, are feed-forward neural networks, as data always move in one direction through the network.

The main layer comprising the CNN is the convolutional layer, which is fundamentally just a set of kernel convolutions acting as feature detectors -- each one aimed at detecting a specific feature in the data. The weights of each cell in the kernel can be adjusted during the training process. Each kernel applies itself across the entire image, and since each kernel is focused on detecting individual features, the outputs of these convolutions are known as feature maps. Due in large part to the shared weights between the multiple feature maps produced by each convolutional layer, CNNs are very tolerant of translations (rotation, movement, scaling, etc...) in their inputs. CNNs will often also contain pooling layers, flattening layers, and fully connected layers. Pooling layers decrease the spatial dimensionality of an input. The max-pooling layer, for example, reduces a layer input of $n\times n$ by to $\frac{n}{p}\times \frac{n}{p}$ by applying a $p\times p$ filter to the input, returning only the maximum value seen by the filter. A flattening layer takes some $n\times n$ input matrix and outputs a length $2n$ vector; this vector can then be input upon fully connected layers as described in Section \ref{ANNsSec}.

Despite implementation difference in CNNs, the same principle of back-propagation is used to adjust the weights associated with each layer. However, because CNNs are often working with data in higher dimensions than traditional ANNs, both forward and back propagation often take longer. Just as in ANNs however, the actual time is heavily dependent on the complexity of the network structure.

For a more in-depth explanation of both ANNs and CNNs see \citet{Geo16}.

\section{Synthetic Data} \label{SD}
Due to the amount of pre-tagged data required to effectively train a neural network, it was not possible to rely on light curves from actual targets. Consequently, we developed an in-house software suite called astroSynth\footnote{https://github.com/tboudreaux/astroSynth} to quickly generate large numbers of synthetic light curves with user-definable parameters, such as pulsation amplitudes and frequencies, noise range, cycle time, visit length, number of visits, average time between visits, and the magnitudes of synthetic targets.

\subsection{astroSynth} \label{SD-AS}
\begin{figure}[h]
	\centering
    \includegraphics[width=0.45\textwidth]{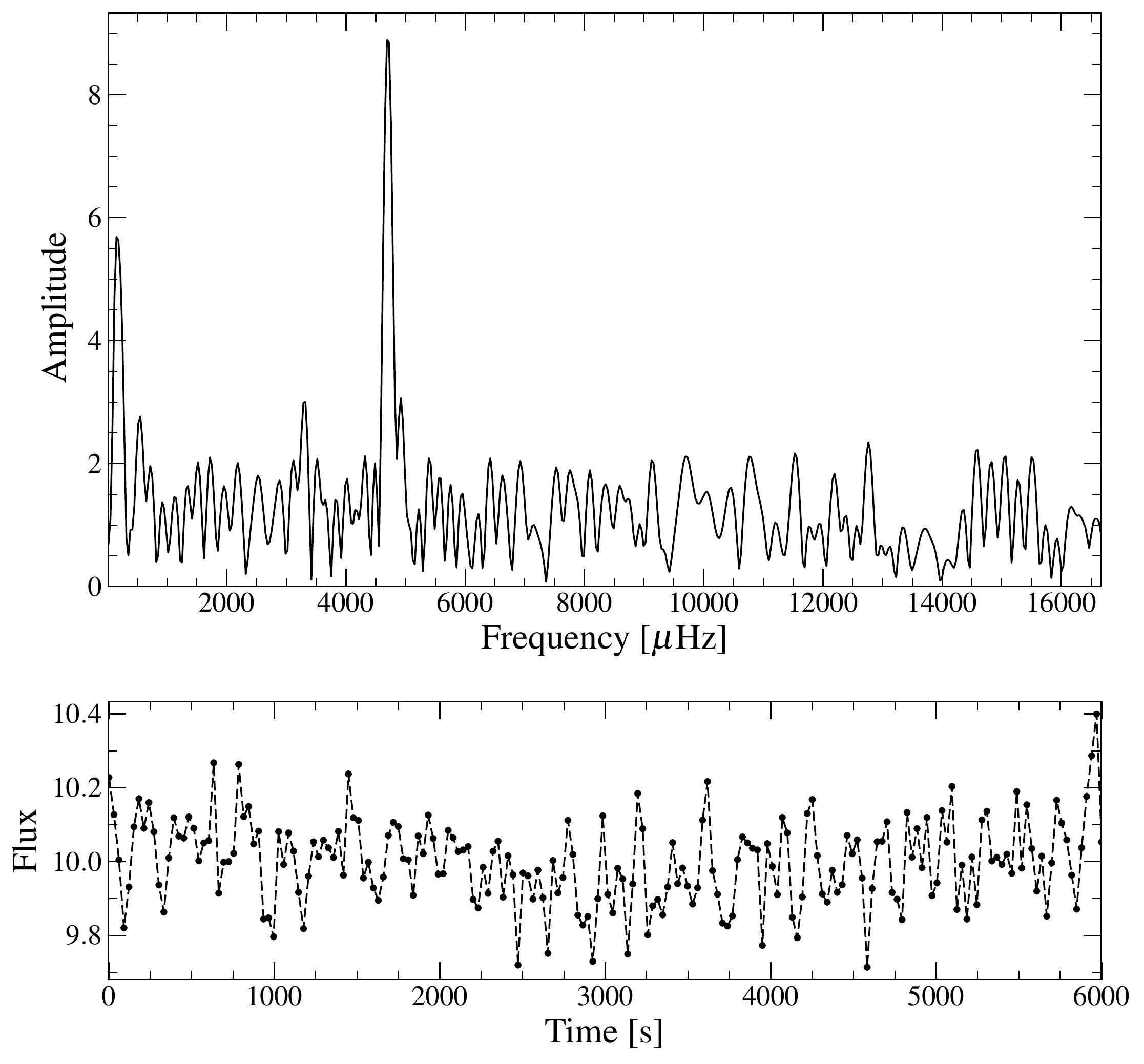}
    \caption{Output from astroSynth.PVS -- generated a single light curve (Bottom). The Lomb-Scargle Periodigram (Top) is accessed via astroSynth.PVS.get\_ft.}
    \label{PVS}
\end{figure}

\begin{figure}[h]
	\centering
    \includegraphics[width=0.45\textwidth]{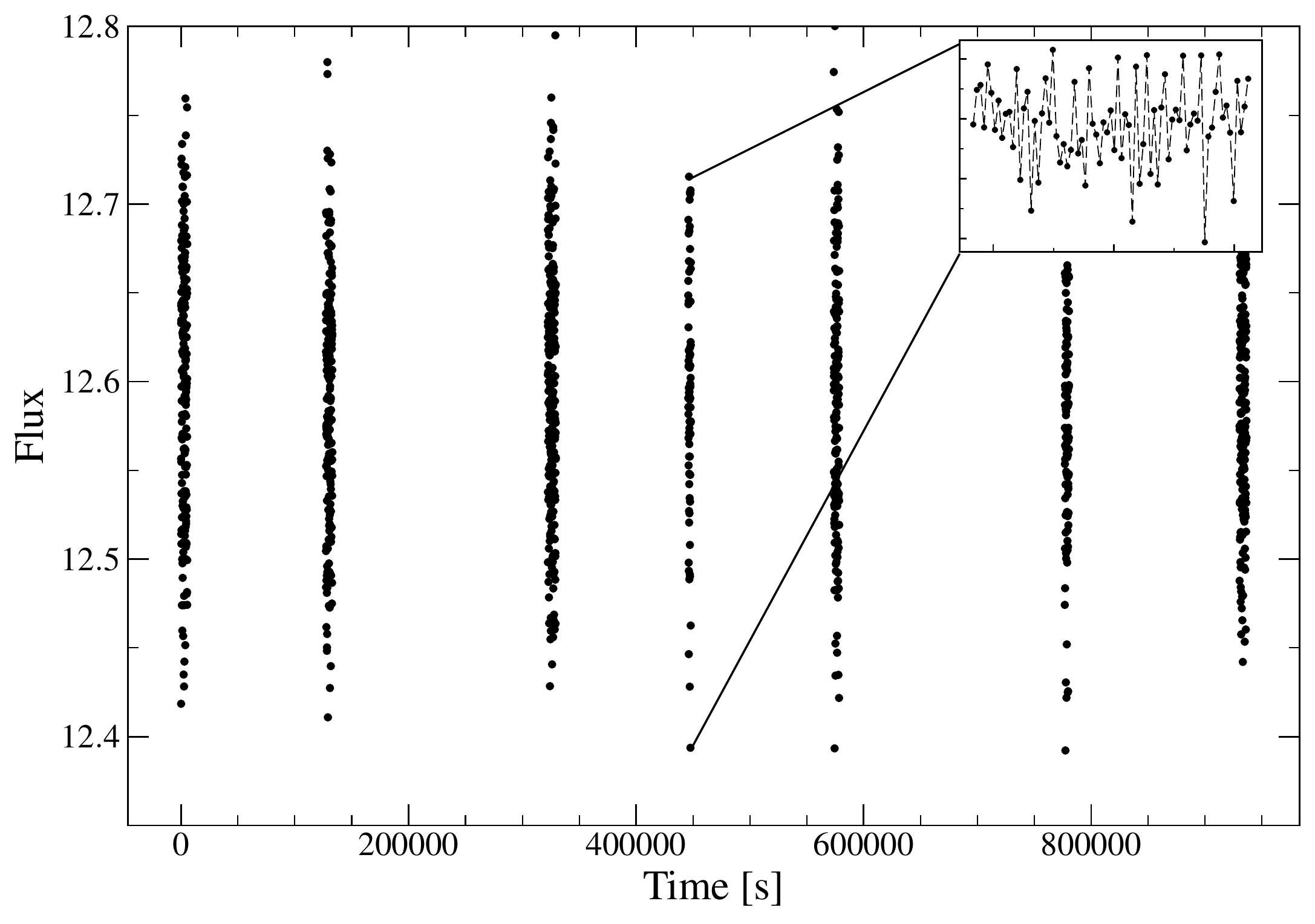}
    \caption{Output from astroSynth.POS -- generated a set of light curves for one target}
    \label{POS}
\end{figure}

astroSynth was developed in Python (3) and it allows for simple function calls to generate large numbers of synthetic light curves. Each light curve is generated by the summation of a set of sine waves and Gaussian noise. We make use of numpy \citep{Wal11} to generate both the sine waves (numpy.sin) and the Gaussian noise (numpy.random.normal). While this is quite a naive method of simulating acoustic mode pulsations, we argue that despite the naivet\'e, the data products of astroSynth can still effectively train a network to classify real data. The reason is that the final structure of the light curve generated in our software is very similar to the structure of an actual light curve. Further, as the main aim of this work is to show that deep learning {\em can} be applied to the classification of pulsating stars, by showing that data of a similar structure can be classified we achieve this goal. In the future when a network aimed at use in actual data-classification problems is constructed, a more physical model of pulsations can be introduced into astroSynth if it proves desirable. An example light curve output from astroSynth can be seen in Figure \ref{PVS}. astroSynth also has the ability to generate non-continuous light curves (Figure \ref{POS}).

Other abilities of astroSynth include: generation of Lomb-Scargle Periodigrams (LSP) from the light curves it produces, dynamic memory management of light curve data -- straddling light curves between memory and disk so orders of magnitude more light curves can be quickly accessed than if they were all stored in memory -- and batch accessing of both light curves and LSPs,  a very useful feature in the training of networks. More information on the abilities and usage of astroSynth can be found on its github page -- https://www.github.com/tboudreaux/astroSynth.

In an attempt to mimic how observations of real stars are conducted, astroSynth generates light curves by first generating an ephemeris for a synthetic target. This ephemeris is defined as the superposition of some number -- from one to the maximum number of desired pulsations modes -- of sine waves. Each sine wave has parameters (frequency, amplitude, phase) chosen from a uniform distribution between the user defined maximum and minimum for that parameter. Poisson noise is then summed into the ephemeris; the noise function is given a centroid at the magnitude of each synthetic target, and the standard deviation of the noise is chosen out of a uniform distribution between a user defined maximum and minimum noise value. Once astroSynth has generated the ephemeris for a synthetic target, an ``observer function'' -- in an analogy to an instrument pointing at a star -- ``looks'' at (records data from) the ephemeris for some time. Light curves returned from astroSynth are the measurements from the observer function. Note that currently astroSynth does not support time evolving pulsation modes. 

\subsection{Our Synthetic Data}
Given that the two network paradigms under investigation are designed for data in different dimensional spaces -- 1D for ANNs and 2D for CNNs -- we elect to generate two separate data sets using astroSynth, each data set will be composed of 100,000 light curves. One of these data sets (hereafter d--I) is composed of continuous light curves, that is light curves without gaps in the observation. The other data set (d--II) is composed of non-continuos light curves, that is light curves which have large time gaps between observations (hereafter referred to having ``multiple visits''). Internally to astroSynth light curves for d--I are produced via the observer function discussed in Section \ref{SD-AS} taking data on d--I's ephemera for their entire length; however, in the case of d--II the observer function will take data from the ephemera, pause, take more data, pause, and so on. 

The properties of the ephemera used in d--I and d--II are the same, except for differences in overall length. In order to keep noise properties comparable between d--I and d--II each observation of the ephemeris must be approximately the same length; as the goal is to have multiple of these visits in d--II separated by large time gaps, the overall length of ephemera used in d--II must then necessarily be longer than those used in d--I. The other ephemeris parameters are defined such that fifty percent of synthetic targets will show properties loosely analogous to those of sdBV$_{r}$ stars (the pulsators), and the remain will be composed of only Poisson noise (non-pulsators). For the pulsators frequencies are allowed to range from 833.3$\mu$Hz to 16670$\mu$Hz, amplitudes from 0 to 20 ppt, and phase from 0 to 2$\pi$. For both pulsators and NOV targets the standard deviation of noise is allowed to range from 1 to 45 ppt.

\section{Artificial Neural Networks Applied} \label{ANNApplied}
\subsection{Synthetic Data} \label{ANNSynth}
The first network paradigm we investigate is the classical fully connected feed forward neural network, the ANN. Intrinsic to many types of Deep Neural Networks (DNNs) -- ANNs included -- is the assumption that inputs will be a constant predefined size, that is, the network will always expect the same number of input parameters. This assumption can be problematic when dealing with light curves, which can vary in length from one observation to the next. There are a few ways in which this input-size problem can be handled, for example: 
\begin{itemize}
 \item Binning light curves into a predefined number of bins.
 \item Running a rolling ``scanner'' of constant size over the data set, passing its reading and a weighted average of the previous zone into the network.
 \item Moving from a time domain to a frequency domain, and in the process defining the number of frequency bins.
\end{itemize}
 
Moving from a time domain to a frequency domain (taking the Fourier Transform of the light curve) was determined to be the most effective strategy, as that transition preserves much of the original information contained within the light curve, while also exaggerating the features that we are most interested in identifying. Additionally it is more easily reproducible by future researchers. The remaining two methods do warrant further investigation. Note however that the rolling scanner method is essentially a very simple Recurrent Neural Network, and as such it would be more productive to investigate the more mature Long Short Term Memory networks (LSTM), a subclass of RNNs, than the version posed in the above list. For more discussion on Recurrent Neural Networks and their possible applications to this problem see Section \ref{Dis}.

An ANN is constructed (hereafter Network A) which expects an input vector with 503 elements. 500 of these are dedicated to the amplitude array of the LSP --- chosen to represent a slight oversampling of the frequency space --- and the remaining 3 are dedicated to the maximum amplitude in LSP, the median value of LSP, and the frequency of maximum value in LSP. While the network could learn these parameters itself, we choose to explicitly include them since it is essentially computationally free to do so, and they are very telling parameters.  Inputs are then passed through an ReLU activation layer, a 20 percent dropout layer, another ReLU activated hidden layer, a final 20 percent dropout layer, and a 2-element softmax activated fully connected layer, read as the network output. Both the standard Keras \textit{adam} optimizer and  \textit{categorical\_crossentropy} loss function were used.

\begin{figure*}
	\centering
    \includegraphics[scale=0.5]{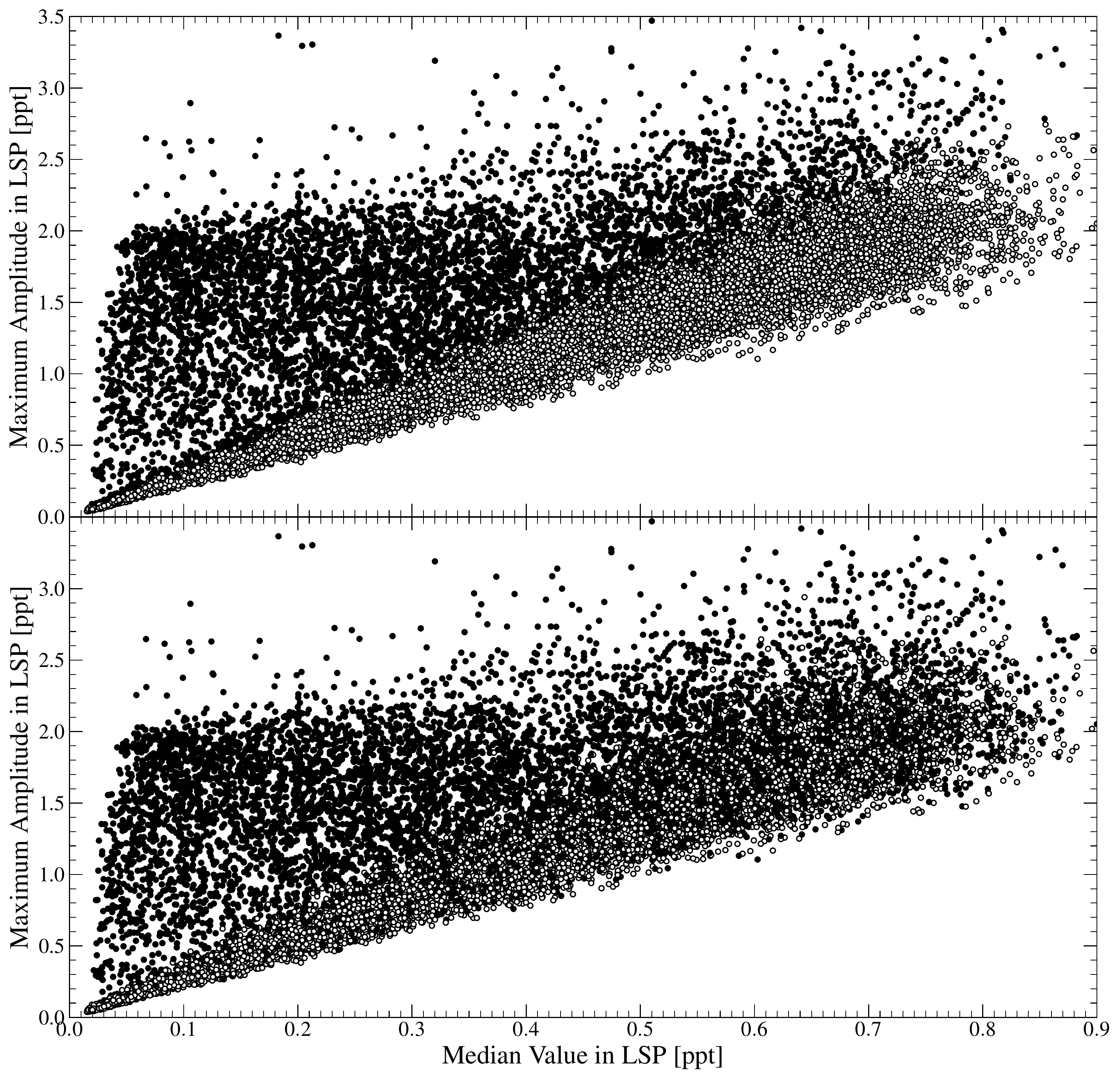}
    \caption{Performance of an ANN visualized in a maximum peak in LSP vs median value in LSP parameter space. (Top) Predicted classification of targets, (Bottom) true classification of targets. The targets which are (or are thought to be) noise are in white, and pulsators are in black.}
    \label{ANNMPMN}
\end{figure*}

Network A is trained using 80 percent of d-I, and validated on the remaining 20 percent. The predictions of the network match to $\sim 95$ percent the true classifications over the entire parameter space (Figure \ref{ANNMPMN}).

\begin{figure}[h]
	\centering
	\includegraphics[width=0.475\textwidth]{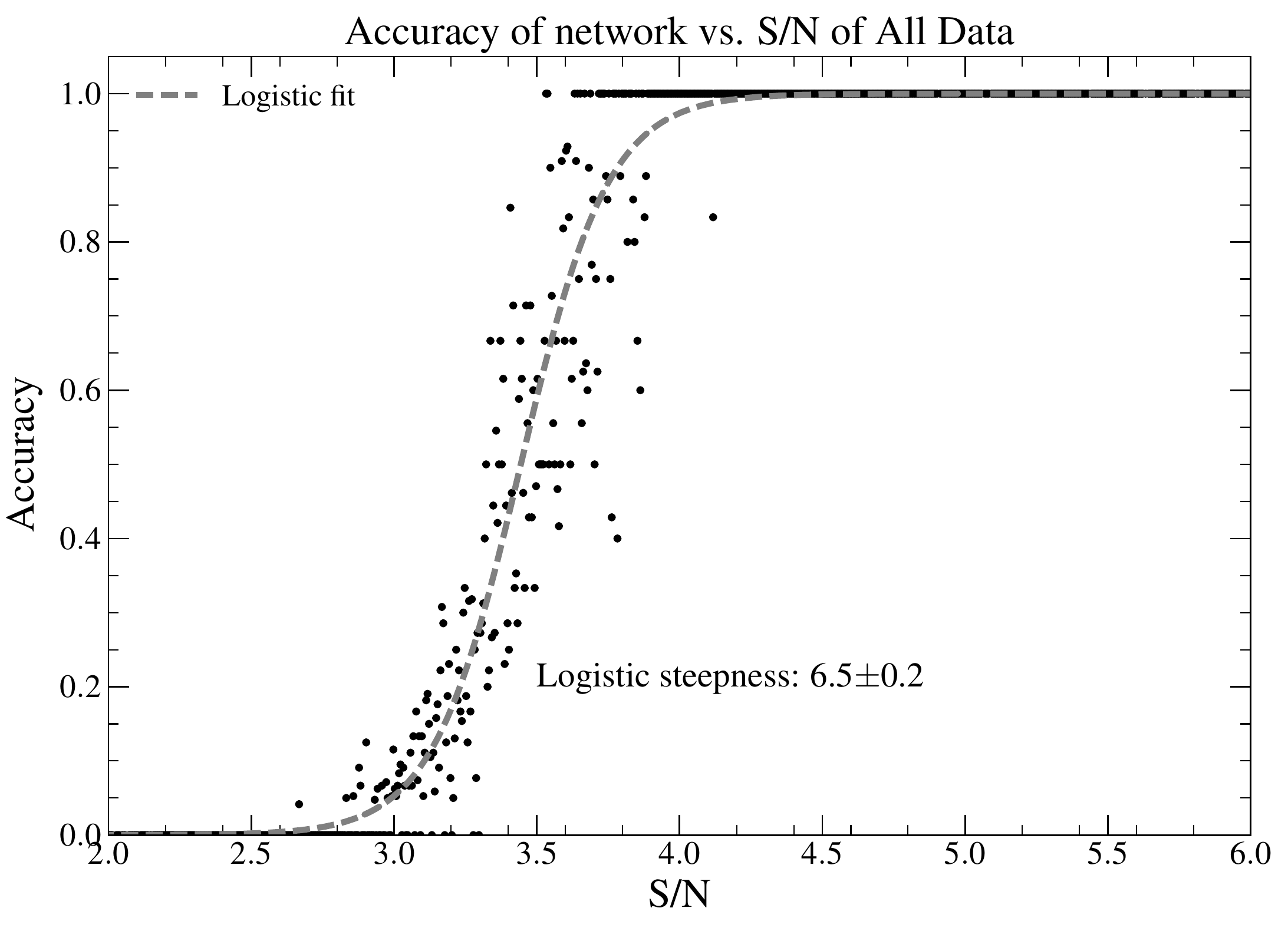}
    \caption{Accuracy of ANN model vs Binned Sigma values in classifying pulsating targets. We can see the networks performance improves drastically around 3.0 sigma.}
    \label{ANNPerf}
\end{figure}

To better understand how the network might perform on real data, we need to understand where and to what extent Network A falters in classification. Figure \ref{ANNMPMN}, while providing a quick method of judging that the network is not outright failing, does a poor job of relaying any quantitative information about how the network performs at different S/N ratios. Instead we bin points together which are $\pm$0.005 sigma of each other, and then calculate the percent accuracy of the predicted classes against the true class for each bin. We see the results of this in Figure \ref{ANNPerf}.

The observed decrease in accuracy at $\sim 3.0\sigma$ in Figure \ref{ANNPerf} is consistent with expectations. As one can see in Figure \ref{ANNMPMN} it is at $\sim3-3.5\sigma$ where the pure noise targets overlap with the pulsators. Using the  standard non-linear curve fitting routines build into SciPy \citep{Jon01} we fit a Logistic function
$$f(x) = \frac{1}{1+e^{-k(x-x_{0})}}$$

\noindent to the accuracy data and estimate the steepness of Network A's change in accuracy to be $k=6.5\pm0.2$, and the offset to be $x_{0}  = 3.444\pm0.006$. Using the accuracy data we then estimate that Network A can achieve an accuracy at or above $\sim90$ percent when classifying signals whose amplitudes are at least $\sim3.44\sigma$ above the noise level.


\subsection{Real Data} \label{RealData}
As telling as synthetic data might be to the classification ability of ANNs, network's abilities to classify real data sets truly tells one whether or not they warrant further investigation. Hence, we feed Network A the light curves of all known sdB stars present in GALEX mission database. As with synthetic data, LSPs are calculated (with 500 frequency bins each) for every visit of all light curves. The amplitudes at each bin -- along with the maximum amplitude, frequency of the maximum amplitude, and median value of the LSP -- are passed to Network A. Given the majority of targets in the catalogue have neither NOV nor pulsator classifications associated with them we are unable to produce either a percent success value, or any such, single number quantifying the overall success of the network. Instead, to get a sense of whether Network A can classify targets we use the five identified pulsators from \citet{Bou17},  we investigate the classification and percent confidence of those classifications in Table \ref{GALEXTargets}.

\begin{table*}
	\centering
	\begin{tabular}{c c c c c}
		\hline
		Target & Visit & S/N & Classification & Confidence \\
		& & & & [\%] \\
		\hline
		\hline
		HS 0815+4243 & 1 & 2.76 & NOV & 78.51 \\
		HS 2201+2610 & 1 & 6.41 & Pulsator & 99.99 \\
		LAMOST J082517.99+113106.3 & 1 & 4.26 & Pulsator & 97.30 \\
		LAMOST J082517.99+113106.3 & 2 & 4.14 & Pulsator & 96.64 \\
		GALEX J08069+1527 & 1 & 7.37 & Pulsator & 100.0 \\
		EC 14026-2647 & 1 & 4.26 & Pulsator & 98.17 \\
		EC 14026-2647 & 2 & 4.74 & Pulsator & 74.17 \\
		\hline
	\end{tabular}
	\caption{Classifications of known $sdBV_{r}$ stars from GALEX data.}
	\label{GALEXTargets}
\end{table*}

Of the five known $sdBV_{r}$ stars, four are successfully identified by Network A, with the remaining target -- HS 0815+4243 -- being incorrectly classed as NOV; however, this is perhaps unsurprising given HS 0815+4243's low S/N -- well below the $3.44\sigma$ line discussed in Section \ref{ANNSynth} (Figure \ref{ANNPerf}). From this we can gleam that Network A, and transitively ANNs in general, can be trained on synthetic data produced with astroSynth to identify rapidly pulsating targets in real data, so long as the signals present in the real data are above $\sim3.5\sigma$. However, we have no way of determining the false-positive identification rate of Network A given the lack of firm classifications for the catalogue. Nonetheless, GALEX's light curves are generally very noisy, which will likely lead to a high false positive rate.

\section{Convolutional Neural Networks Applied}
\begin{figure}[h]
	\centering
	\includegraphics[scale=0.5]{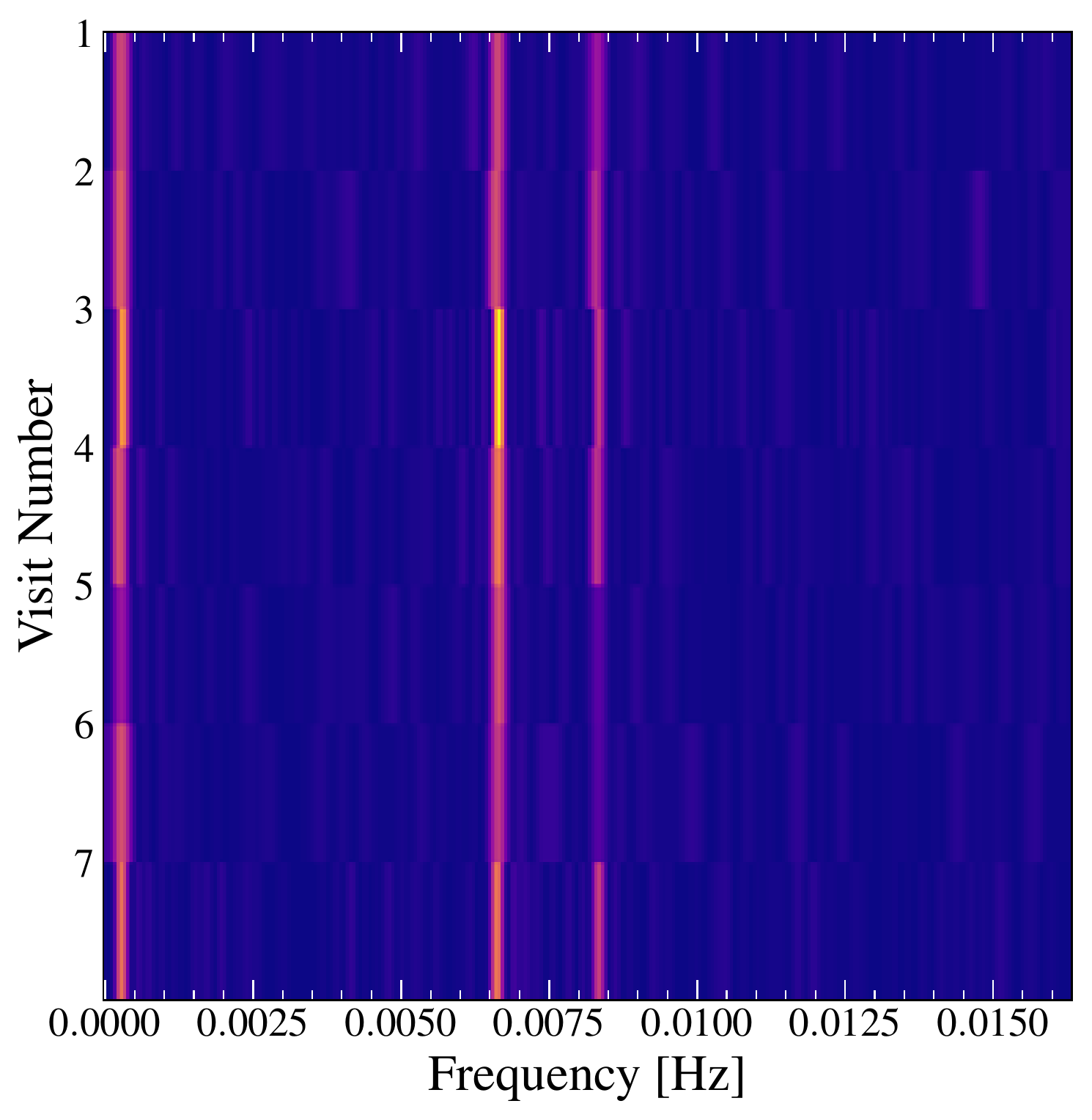}
	\caption{sliding FT showing 3 modes of pulsation. Note that pulsation amplitudes have been exaggerated in this figure to highlight their existence.}
	\label{SFT}
\end{figure}

Given the success we found using ANNs with d-I, we wanted to make our data more physical. As discussed in Section \ref{SD} to accomplish this we modeled the multiple visits that researchers generally have on an object. d-II has non-continuous light curves for each target, which can have visits separated by large amounts of time. As such we elected to take the LSP of each visit individually, as opposed to the LSP of the entire light curve. These LSPs are taken through time, so by stacking them into a 2D array where the value at each index is amplitude, setting the vertical axis as time, and the horizontal as frequency we can generate a ``sliding FT'' (Figure \ref{SFT}). 

Sliding FTs are already in the form of an image; therefore, it makes sense to use CNNs for analysis. Before we pass the sliding FTs into a network we apply some basic transformations to it which will allow the CNN to learn its features more easily (note that when performing any analysis using a CNN, the same transformations should usually be applied). First we scale all values so that they fall between 0 and 1, inclusive. Then all sliding FTs are reshaped into a square. Reshaping is achieved by stretching each individual LSP over multiple rows until the total number of rows is equal to the number of frequency bins, which we fix at 300 --- a slight undersampling of the frequency space which was made to decrease runtime, as the complexity of Network B scales like the square of the number of frequency bins used. Stretching is performed using a combination of two methods: one, take the desired height of the image and divide that by the number of visits. Floor the resulting value, then duplicate each visits's LSP by the result of that floor operation. Two, scipy.misc's resize function, this is applied only after the previously described stretching operation and handles cases where the desired vertical dimension cannot be achieved with an integer multiple of the number of visits. Very little interpolation should have to be done as, however when and where it is required the image resize function will use cubic spline interpolation. Here we initially ran into the issue that scipy's imresize function also rescales all values in the 2D array being resized to between 0 and 255. We undo this rescaling, however as the rescale operation rounds all of its values and then casts them into integers undoing the operation introduces more noise, this additional noise however is on average $\sim0.05$ppt, well below the noise level of any given target, and as such should not significantly alter any results.

\begin{figure*}
  \centering
  \includegraphics[scale=0.5]{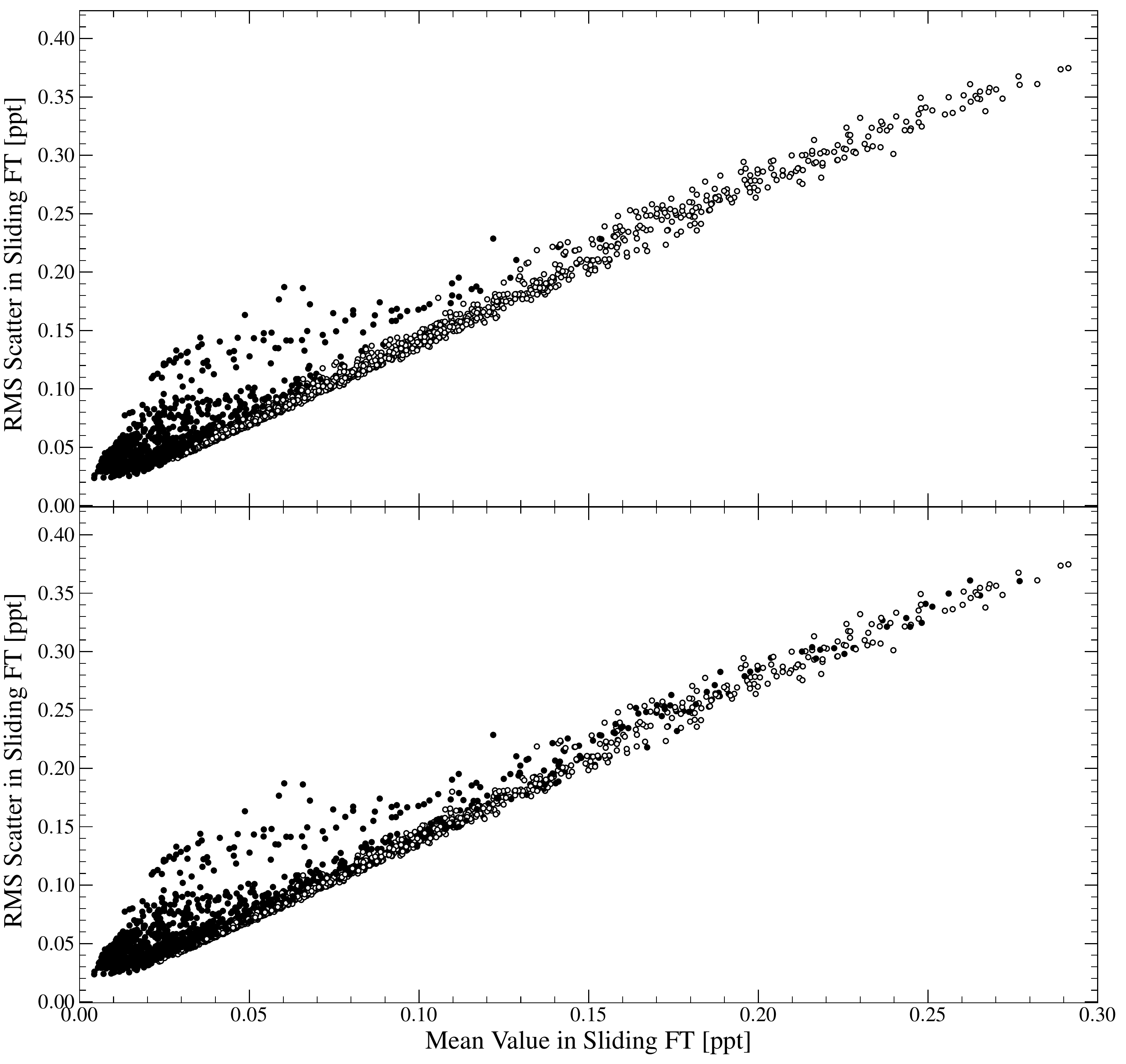}
  \caption{CNN Classification of 2000 targets' sliding FT (Top). True Classification of 2000 targets' sliding FTs (Bottom). RMS is calculated as 1/median value in sliding FT and the max value in all sliding FTs has been normalized to 1.}
  \label{CNNClass}
\end{figure*}

A network is constructed that expects an input of a 300 x 300 matrix with one channel per data entry (hereafter Network B). A convolutional layer then makes use of a 3x3 kernel to generate 32 feature maps. These are activated with a ReLU, passed through a 20 percent dropout layer, and then a max pooling layer with a 2x2 kernel (thus reducing the overall size of the image by a factor of 4). The outputs from the max pooling layer are flattened (i.e. 10x10 matrix would become a length 100 vector), passed to a fully connected layer, activated with an ReLU, then to a 30 percent dropout layer, and finally a two-element output layer activated with a softmax function. We use the standard keras optimizer ``adam'', and calculate loss using the standard keras ``categorical\_crossentropy'' loss function.

d-II consists of 100,000 targets, each with between 1 and 50 visits and with all other properties (pulsation amplitude range, frequency range, etc...) the same as in d-I. Eighty percent of d-II is used as the training dataset, with the remaining twenty percent used for validation. Figure \ref{CNNClass} shows 2,000 targets plotted (due to memory limitations of the host computer used for this work we are unable to plot all 20,000 targets used for model validation) in an RMS scatter vs. mean value in sliding FT parameter space, and manages to show the separation between pulsators and non-pulsators quite well. As expected, Network B performs well where the pure noise targets and pulsators barely coexist. It performs more poorly in the opposite case. In fact, upon initial investigation of Figure \ref{CNNClass} it seems that Network B performs comparably to Network A. This model does, however, appear to over-classify pulsators as there are far more false pulsators present than there are false noisy targets. 

\begin{figure}[h]
	\centering
	\includegraphics[width=0.475\textwidth]{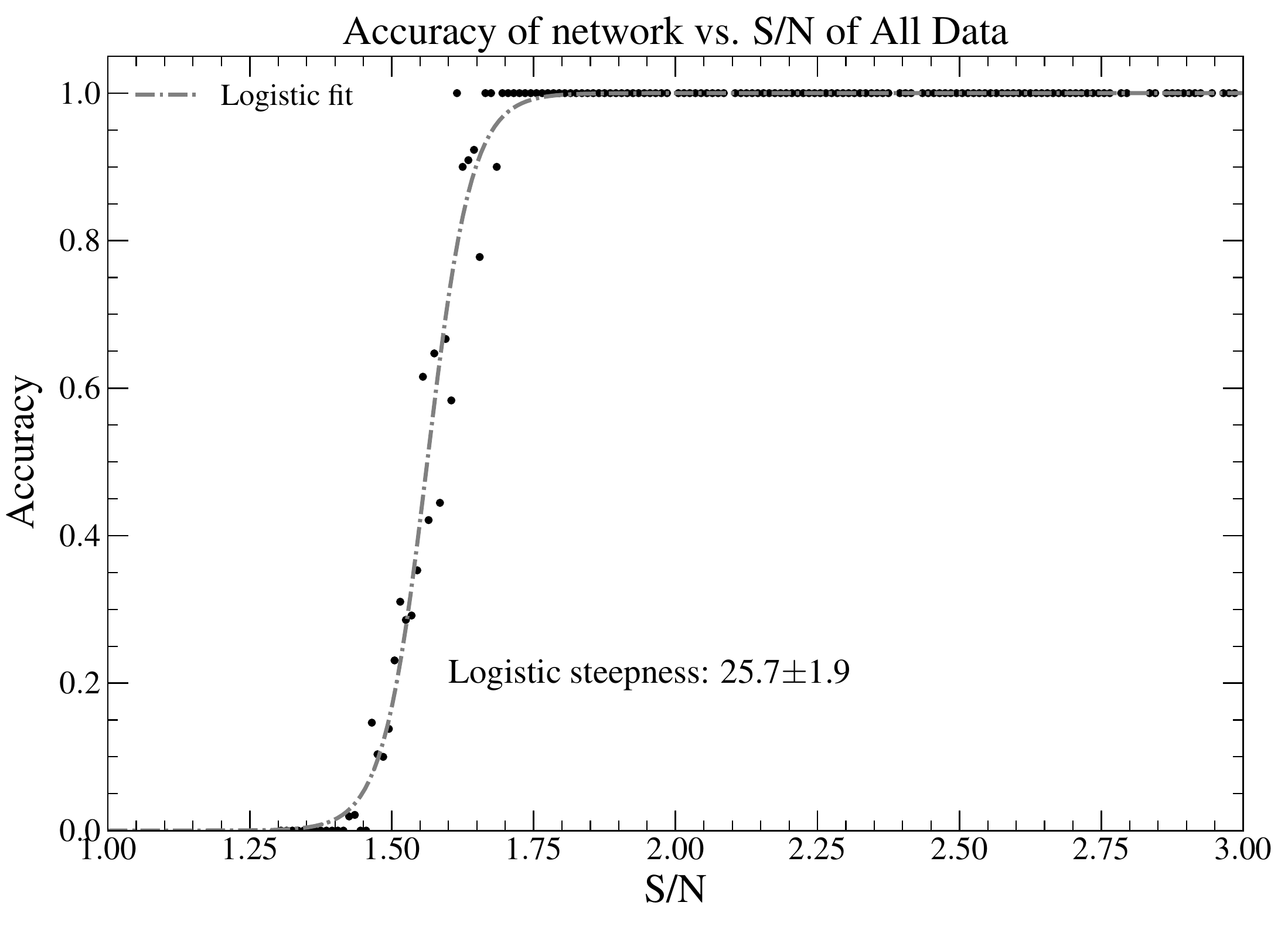}
  \caption{Performance of CNN visualized as accuracy in classification of pulsating targets of model classification vs signal to noise in data. Note that due to memory limitations on the host computer of this work we were unable to plot all 20,000 points used to validate Network B, here we show the first 2,000 points of those 20,000}
  \label{CNNPerf}
\end{figure}

We use the same method to rigorously quantify Network B's performance as was used in Section \ref{ANNSynth}; that is we investigate model accuracy vs signal to noise level in the sliding FT (Figure \ref{CNNPerf}). Figure \ref{CNNPerf} allows us to clearly see the improved performance of network B's analysis on d--II over network A's analysis of d--I. Using a non-linear least squares fitting routine we again fit a logistic sigmoid function to the accuracy vs S/N data. This fit has a steepness of $k=25.7\pm1.9$, and a centroid $x_{0} = 1.563\pm0.003$. Finally we use the fit to estimate that Network B can achieve 90 percent or greater accuracy when the target signal's amplitude is at or above $1.56\sigma$.

We do not attempt to apply Network B or any CNN to the GALEX data set as we did in Section \ref{RealData} due to the small number of visits known pulsators have.

\section{Discussion} \label{Dis}
Deep-learning offers an enticing method of data analysis. Given its promise of complex-feature detection in minimally prepared data, one would imagine that researchers would flock to use these algorithms. However, because of the difficulty in finding enough tagged data in the correct form, the adoption of not only deep learning but many machine learning algorithms in astronomy has been curbed. It is difficult to impossible to find enough tagged data in the correct form for any given problem to effectively train these algorithms. We handle this problem by generating synthetic data, which despite the naivet\'e of the data-generation model, matches the overall structure of real p-mode pulsator light curves quite well. In the future steps will be taken to better match the synthetic data generation model to physical observations. These improvements will take the form both of accounting for more complex physics such as rotational splitting, as well as better matching the amplitude and frequency distribution of sdBV$_{r}$ stars. This last point is important. Currently we can only make statements about the effectiveness of our networks down to certain sigma or S/N levels, not what percentage of actual sdBV$_{r}$ stars would be successfully identified. When we match the distribution we will be able to make an approximation of the latter statement. We also recognize that by focusing solely on acoustic mode pulsators we have ignored other types of sdB variability. This choice to focus on sdBV$_{r}$ stars was made due to time constraints, and a desire to limit the scope of initial investigations; however, given the success we have found here, future work will analyze both gravity mode pulsations and eclipsing binaries. 

When interpreting the results presented here it is important to note that very little in the way of tuning the network's structural elements was done. Such elements, known as hyperparameters include the number of layers, how deep each layer is, the learning rate $\eta$, etc.... They {\em can} have a significant effect on a network's performance. It is therefore conceivable, and in fact likely, that with careful tuning the networks presented here could be outperformed. The standard method of tuning hyperparameters is to build an n-dimensional grid of the parameters, try every possible network configuration, and use the most effective one. Due to the expensive nature of this tuning, and the fact that we found good results without dedicating a large amount of time to it, we elected not to do this work. In the future, when these networks are being aimed towards an analysis pipeline, hyperparameter tuning should certainly be carried out. 

Working specifically with time-domain data posed a problem because while the network expected an input vector of a certain, pre-defined size, the data set could very well be, and most often would be, a different size. Here this issue was handled by moving from a time domain into a frequency domain -- the number of frequency bins being the size input expected by the network (or in the case of the CNN the same thing but the dimensionality of the sliding FT being the dimensionality expected by the network); it would, in the future, be interesting to investigate the ideal ratio of frequency bins to frequency resolution. While this method provided promising results, it would be interesting to see how a network would perform if it learned from the time-series data directly, as certain features are lost or hidden when moving in frequency space. For example, in a target with multiple visits a network might be able to correlate phase information related to the pulsations between light curves, however by moving out of time space we loose phase information and that route is now closed off. Analysis routes such as these could open the door for signals at or below the noise level to be effectively identified.

The other main advantage to staying in time space is that the processing of the data is significantly reduced. LSPs calculated here use scipy's Lomb-Scargle method which goes like $O(n^{2})$, and even when using the fast LSP method built into astropy (which goes like $O(n\ln{n})$), generating the LSPs was by far the most time intensive part of this work (including training the networks). If this could be cut out in favor of directly learning from the time series data then significant amounts of time would be saved. Finally one must consider the value of being able to analyze time series data not as well suited to Fourier transforms as light curves from a pulsating star might be, such as an eclipsing system, or cataclysmic variables. 

Recurrent Neural Networks (RNN) are able to analyze, and in fact are well suited for analysis of, time-series data regardless of length variations. A discussion of how RNNs work is well beyond the scope of this paper. It is enough to know that recurrent neural networks share state through time, that is to say, that RNNs have memory, and can change their decisions based on things they have seen in the past. Note that this process is separate from training the network. No weights are being modified; rather, a value is being continually passed from the output of layers back into those same layers. What this allows for is analysis of data of an arbitrary length by sliding a window over it, and reading the network output only when the window has passed over the entire dataset. RNNs, and specifically a subtype called Long Short Term Memory networks (LSTM) are being widely used in time series forecasting, and applications relating to Natural Language Processing --- it should be noted however, that the training process for an RNN is often significantly more time intensive than for either a CNN or ANN. {\em In the future, work which aims to analyze time series data should focus on the use of RNNs} as they are specifically designed to handle such problems quickly and efficiently.

We must understand that the performance of the networks presented here is a function of the data they were trained with. This may seem obvious; however, the effects of this run deeper than just being able to identify pulsations within the range of amplitudes and frequencies used when generating the data. Rather, the effect of using this data set is that any patterns that may be present in real data will be unknown to the network. For example, there is no weight placed on certain frequencies over others as the frequency range is uniform. No thought is given to these patterns because the network has never encountered them. What we have essentially presented here is a worst-case scenario. So while in the future it is important that we emulate any such patterns that may exist it seems unlikely that they would deteriorate the performance of any network. Instead they would, at the least, not affect the performance, and possibly help the network improve. We also recognize certain limitations of the data model used here. For example by taking the LSP of each visit individually, no signal longer than the observational cadence can be measured. Acoustic mode pulsations are unlikely to be lost due to this effect (due to their short periods); however, if rotational splitting had been modeled this issue --- with losing signals --- may have been more pronounced depending on the periodicity which lead to the rotational splitting.

Finally we would urge future researchers to not fall into the trap of overestimating the abilities of deep learning. Deep learning in so far as it is applied to astrophysical research is a field in its infancy, and it is both easy and tempting to imagine a future where a multitude of problems are solved via deep learning. Maybe this will be the case; however, like any other method that claims its roots in some form of scientific rigor, deep learning presents a single possible model. This model is one that is, at least currently, generated by a relatively enigmatic black--box, namely the hidden layers of a network; and consequently, one should {\em always} follow up any statement made by a deep neural network with an in depth case study, and not rely solely on the judgment of a set of matrix multiplications.

\section{Conclusion}
Using two kinds of deep learning algorithms we show that sdBV$_{r}$ pulsators whose modes of pulsations are both visible in the frequency domain to the human eye and above the noise level can be identified quite well, at an accuracy of 90 percent down to $\sim3.6\sigma$ with ANNs, and down to $\sim1.6\sigma$ with CNNs. Both the more traditional fully connected, or Artificial Neural Networks, and the image-focused Convolutional Neural Networks, perform well here, however our CNN (Network B) is able to identify signals at a lower signal to noise than our ANN (Network A) is able to. We conclude that these are effective means of analyzing medium to high signal to noise pulsators, but that careful tuning of network hyperparemeters is likely necessary if one wants to extract the full potential of a network. Finally, future work should focus on the use of Recurrent Neural Networks to analyze data in a time domain as opposed to analyzing in a frequency domain, as we were essentially forced to do here.
 
\section*{Acknowledgements}
This research has made use of NASA's Astrophysical Data System (ADS). I would not have been able to conduct this work without the unwavering support of my wonderful adviser; thank you Brad. This research was also made possible by the generous support of the High Point University (HPU) Student Government Association (SGA), the HPU Department of Physics, and the HPU department of Math \& Computer Science. I would like to thank the SOC and the LOC of sdOB8 for organizing the conference where this work was originally presented.

\end{document}